\documentclass[final,5p,times,twocolumn]{elsarticle}
 \biboptions{comma,sort&compress}
\usepackage{graphicx}
\usepackage{here}
\usepackage[dvips]{epsfig}
\usepackage{cuted}
\def\beq{\begin{equation}}
\def\eeq{\end{equation}}
\def\bea{\begin{eqnarray}}
\def\eea{\end{eqnarray}}
\def\bq{\begin{quote}}
\def\eq{\end{quote}}

\def\nnb{\nonumber}

\def\nnb{\nonumber}

\def\la{\langle}
\def\ra{\rangle}

\def\ba{\vspace*{-0.2cm}\begin{array}}
\def\ea{\end{array}\vspace*{-0.2cm}}

\def\b{$\bullet~$}

\def\als{\alpha_s}

\def\gg2{ \la\alpha_s G^2 \ra}
\def\gg3{g^3f_{abc}\la G^aG^bG^c \ra}
\def\ggg4{\la\als^2G^4\ra}

\begin{document}
\begin{frontmatter}

\title{The First Months of COVID-19 in Madagascar} 
 \author[label3,label2]{Stephan Narison}

\address[label3]{Laboratoire
Particules et Univers de Montpellier, CNRS-IN2P3, 
Case 070, Place Eug\`ene
Bataillon, 34095 - Montpellier, France.}
 \address[label2]{iHEPMAD Research Institute, University of Ankatso, Antananarivo 101, Madagascar.}
   \ead{snarison@yahoo.fr}

\markright{The First Months of COVID-19 in Madagascar}
\begin{abstract}
\noindent
Using the officially published data and aware of the uncertain  source and insufficient number of samples, we present a first and (for the moment) unique attempt  to study the first two months spread of the pandemic COVID-19 in Madagascar.
The approach has been tested by predicting the number of contaminated persons for the next week after fitting the inputs data collected within 7 or 15 days using standard least $\chi^2$-fit method. Encouraged by this first test, we study systematically during 67 days , 1-2 weeks new data and predict the contaminated persons for the coming week. We find that the first month  data are well described by a linear or quadratic polynomial with an increase of about (4-5) infected persons per day.  Pursuing the analysis, one note that data until 46 days favours a cubic polynomial behaviour which signals an eventual  near future  stronger growth as confirmed by the new data on the 48th day. We complete the analysis until 67 days and find that the data until 77 days confirm the cubic polynomial behaviour which is a remarkable feature of the pandemic spread in Madagascar. We expect that these results will be useful for some new model buildings.  A comparison with some other SI-like models predictions is done. 
These results for infected  persons may also be interpreted as the lowest values of the real cases due to the insufficient number of samples (about 12907 for 27 million habitants on 05/06/20). The data analysis of the absolute number of cured persons until 67 days shows an approximate linear behaviour with about 3 cured persons per day. However, the number of percentage number of cured persons decreases above 42-46 days indicating the limits of the hospital equipment and care to face the 2nd phase of the pandemic for the 67th first days. Some comments on the social, economical and political impacts of COVID-19 and confinement for Madagascar and, in general, for Worldwide  are shortly discussed.
\end{abstract}
\end{frontmatter}
 \vspace*{-0.25cm}
 \section{Introduction}

 COVID-19 is scientifically named SRAS (Syndrome Respiraroire Aigu Severe) or SARS (Severe Acute Respiratory Syndrome)-CoV-2 or COVID-2-19 or COVID-19 in reference to the SRAS-CoV or Coronavirus pandemic (2002-2003) starting  in China and propagated in 30 countries which has affected 8 000 persons and caused 774 deceased\,\cite{PASTEUR}.  There is a slight difference between the two COVID in the genome structure of the virus thus the symbol 2. 

   Another pandemic named MERS-CoV (2012-2013) or SRAS of the MIDDLE EAST has been found for the first time in Saudi Arabia and has affected 1 589 persons and caused 567 deceased in 26 countries.

 The CORONAVIRUS are of animal origins. An asymptomatic animal  transmits to another ones and then to human. It was bats for SRAS-COV and MERS-COV which transmit the virus to masked palm civets for China and to camels for Saudi  Arabia as intermediates and then to humans. 

 COVID-19 has been supposed to start from  the Wuhan local market (Hubei province - China) in December 2019.  It may also come from bats which have transmitted the virus to the pangolin sold in the Wuhan market and then the pangolin has transmitted to human. However, the origin of the transmission is not quite clear and there are some hypotheses that the virus may have been invented inside the Wuhan research laboratory and has been propagated inadvertently. 
 The transmission mode from humans to  humans is through saliva droplets, sneezes, coughing and contacts.

The virus has subsequently propagated to the rest of China and contaminated up to now (15/04/2020) 82 295 persons. Today, almost all countries of the  world are now affected leading to 1 997 321 infected persons with 500 819 cured and 127 601 deceased. The 5 countries with most affected persons relative  to the total number of cases and to the population density are shown in Tables\,\ref{tab:countries} and\,\ref{tab:countries2}, where one can notice that the relative number per million of infected persons is very high in small area countries.

COVID-19 then appears as  the most devastating pandemic of the 21th century\,\footnote{For a more complete but simple review on the coronavirus, see e.g.\,\cite{SANTE}.}. 
 To face this unexpected drama where there is no known medicine\,\footnote{Drugs based on chloroquine proposed by Pr. Didier Raoult - Marseille (expert in tropical diseases), which were successful against paludism are promising though under debates. Some other pists are the uses of some endemic medicinal plants such as in Madagascar where some of them contains natural chloroquine.} and vaccine\,\footnote{There is some proposal from the Bill Gates foundation but doing the first tests in Africa are criticized.} to fight the virus, only urgent preventive measures have been taken by different countries. 
  
  As the virus is quite heavy, it cannot travel more than 1 meter\,\footnote{However, new study  (French TV source) recommends a distance above 2m.}. Its lifetime is relatively short on skin (few minutes) but long in air and copper (3-4) hours, clothes (12 hours),  cartons (24 hours), woods, plastics and metals (3-9 days)\,\footnote{The true lifetime of the virus on inert objects is still under study and needs to be confirmed.}. So different  preventive barriers have to be used (masks, soaps, hydroalcoholic gels,..) to stop the virus.

However, in addition to these barrier precautions,  drastic measures such as confinement have been taken by governments of different countries which ineluctably affect the social organizations and economic situations. Confinement may lead to a new form of society and economy, to a new way of leaving  in the near future. 

Different analysis of the propagation of COVID-19 appearing in China, Europe, UK, USA and Russia have been done in different works (see e.g. various new articles in arXiv\,\cite{ARXIV}) where most of them are based on Susceptible-Infected (SI) or Susceptible-Infected Recovered (SIR)\,\cite{SIR} models (see e.g.\,\cite{ZIFF,MALT,MLI,ZLIU,MO,SARDAR,SAVI,CARCIONE,HERY,HERY2,GRECE1,GRECE2}) while some is a simple Gaussian fit of the data\,\cite{TSIRONIS}.

To our knowledge nothing has been yet done in African and some other developing countries. The reasons could be that, in these developing countries,  the pandemic is still at the beginning of its effect and the data are not yet sufficient to make a rigourous statistical study (Gaussian law of large numbers of events) which is also complemented by the eventual non-reliability of the collected data due to the few numbers of detection tests or perhaps for some political reasons. 
\section{The spread of infected persons}
\subsection{First analysis : data up to 15 days}

COVID-19 enters in Madagascar on 21th March through the Antananarivo-Ivato airport via passengers from Europe and from China. The official measure for closing the
airport was too late despite the requests of many individuals or through the social networks. 
 
We collect, in Table\,\ref{tab:data}, data from different national newspapers\,\cite{MIDI} and TV\,\cite{TVM} issued from  the national representant \,\cite{CCO} of the World Health Organization (WHO)  / Organisation Mondiale de la Sant\'e (OMS). 

 We use the data of the first 15 days to predict the ones from 15 to 20 days. Noting that some data do not agree each others and that the detection tests are only done for few samples of the population, we have introduced some estimated errors  in order to quantify this deficit. 

\begin{figure}[hbt]
\vspace*{-0.25cm}
\begin{center}
\includegraphics[width=8.5cm]{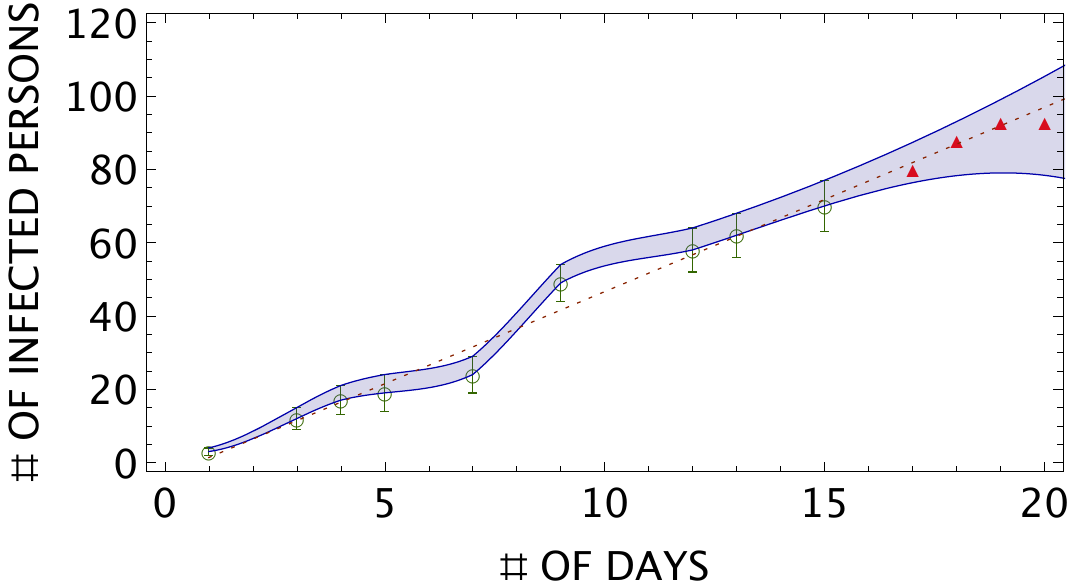}
\vspace*{-0.25cm}
\caption{\footnotesize  Number of new contaminated persons per day versus the number of days : green circles are the input data, filled region are the predictions limited by the lower curve (central value of the data) and by the upper curve (data with  added positive estimated errors). Dashed red line is the linear fit. Triangles are data received after the fitting procedure.} 
\label{fig:fig1}
\end{center}
\vspace*{-0.5cm}
\end{figure}

 The analysis is shown in Fig.\,\ref{fig:fig1} where the input data are the green circles.  Our prediction (filled region) comes from a standard least-$\chi^2$ fit of the data using a Mathematica Program elaborated by\,\cite{MATH}. The  central value of the data is fitted by the lowest curve while the central value $\oplus$ the estimated errors by the upper curve. The fit is extrapolated until 20 days. 

 The new data points in red are inside our prediction region which we consider as an encouraging positive test of the approach based on a standard least-$\chi^2$ fit of the data.

We also show (dotted curve)  the expectation from a simple linear parametrization of the data as given by\,:
\vspace*{-0.2cm}
\beq
n_i(x)\simeq -3.67 +5.03x~,
\label{eq:linear1}
\eeq
\vspace*{-0.2cm}
where $n_i(x)$ is the number of daily new infected persons as a function of the number $x$ of days. The data show a linear increase for about 5 persons per day. 
\vspace*{-0.25cm}
\subsection{Second analysis : data up to 30 days}
\begin{figure}[hbt]
\vspace*{-0.25cm}
\begin{center}
\includegraphics[width=8.cm]{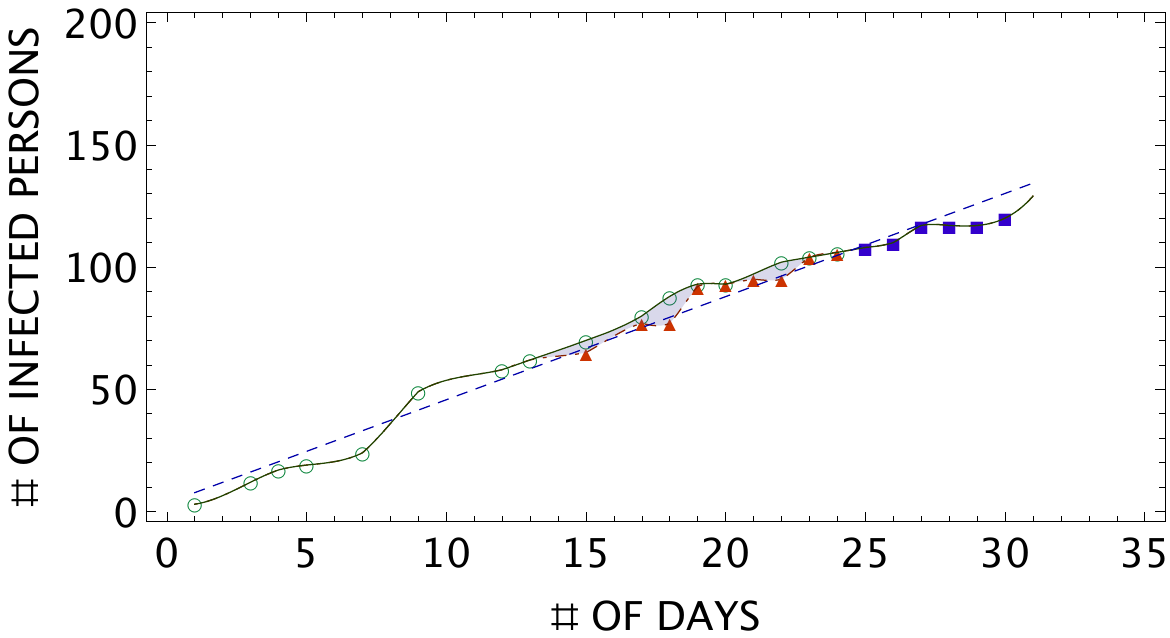}
\vspace*{-0.25cm}
\caption{\footnotesize  Number of new daily infected persons versus the number of days : green circles are the input data from\,\cite{MIDI,GOOGLE}. The red triangles are data from\,\cite{OMS} from the 17th day. Blue rectangles are new data\,\cite{MIDI,GOOGLE,OMS}  from 25th to 30th days. The continuous oliva  curve is the fit from 1 to 30th days of\,\cite{MIDI,GOOGLE} data while the red dot-dashed one is the fit of\,\cite{OMS} data which follows the red triangles data points. The dashed blue line is the linear fit given in Eq.\,\ref{eq:linear2}. } 
\label{fig:fig3}
\end{center}
\vspace*{-1cm}
\end{figure}

 Encouraged by this positive test, we improve the analysis by adding the new data (see Fig.\,\ref{fig:fig3}) collected from the 16th to 30th days. 

 The $\chi^2$-fit is shown by the continuous green line (green circle data from\,\cite{GOOGLE,MIDI} and dot-dashed red curve (red triangle data from\,\cite{OMS}).

Doing a linear fit of the data collected during 24 days, we obtain :
\vspace*{-0.2cm}
\beq
n_i(x)\simeq -1.42 +4.73x~,
\label{eq:linear2}
\eeq
\vspace*{-0.2cm}
while adding the data until 30 days (blue rectangles in Fig.\,\ref{fig:fig3}), we obtain :
\vspace*{-0.2cm}
\beq
n_i(x)\simeq 3.5 +4.22x~.
\label{eq:linear3}
\eeq
\vspace*{-0.2cm}
The slopes of the linear curves obtained in Eqs.\,\ref{eq:linear2} and\,\ref{eq:linear3} ((blue dotted line) are consistent with the one in Eq.\,\ref{eq:linear1} shown in Fig.\,\ref{fig:fig3}. This linear fit can be improved by using a quadratic fit of the data. In this way, one obtains :
\vspace*{-0.2cm}
\beq
n_i(x)\simeq -8.14 +6.45x-0.07x^2~.
\label{eq:quadratic}
\eeq
\vspace*{-0.2cm}
The fits indicate an average of about (4-5) persons infected per day. 
\subsection{Comments on these results}
The previous analyses have been done using data from\,\cite{GOOGLE,MIDI,OMS}. We are aware that these data may be inaccurate as there are some periods (week-end) where the data remain stable which may due to the break of the infection tests service.  These data may also underestimate the real number of new daily infected persons due to the insufficient number of detection tests for the whole country which contains about 80\% of rural population (see e.g\cite{SNBm}). The recent news\,\cite{MIDI} indicates that about 2200 tests have been performed. The detection tests should be extended for improving the quality and reability of the analysis.

However, despite these warnings, the reported data can be considered as lower bounds of the real case and the results from the previous study may already be useful. The results of the analysis can be an important guide to the  government for taking the right decision at the right time in order to control this tremendous pandemic. 

 Our analysis concerns only the officially declared contaminated cases and does not subtract the numbers of cured and deceased persons. 

 However with the last collected number (19/04/2020) for about 2 200 tested persons\,\cite{MIDI} which are relatively small compared to the 27 million of population, one has\,\cite{GOOGLE,OMS} 120 declared cases, 35 cured persons and 0 deceased.  One can also notice that the number of cured persons increases linearly for about 2 persons per day.

\subsection{Comparison with some other approaches}
\begin{figure}[hbt]
\vspace*{-0.25cm}$\overline{\eta}_c\eta_c$
\begin{center}
\includegraphics[width=8.cm]{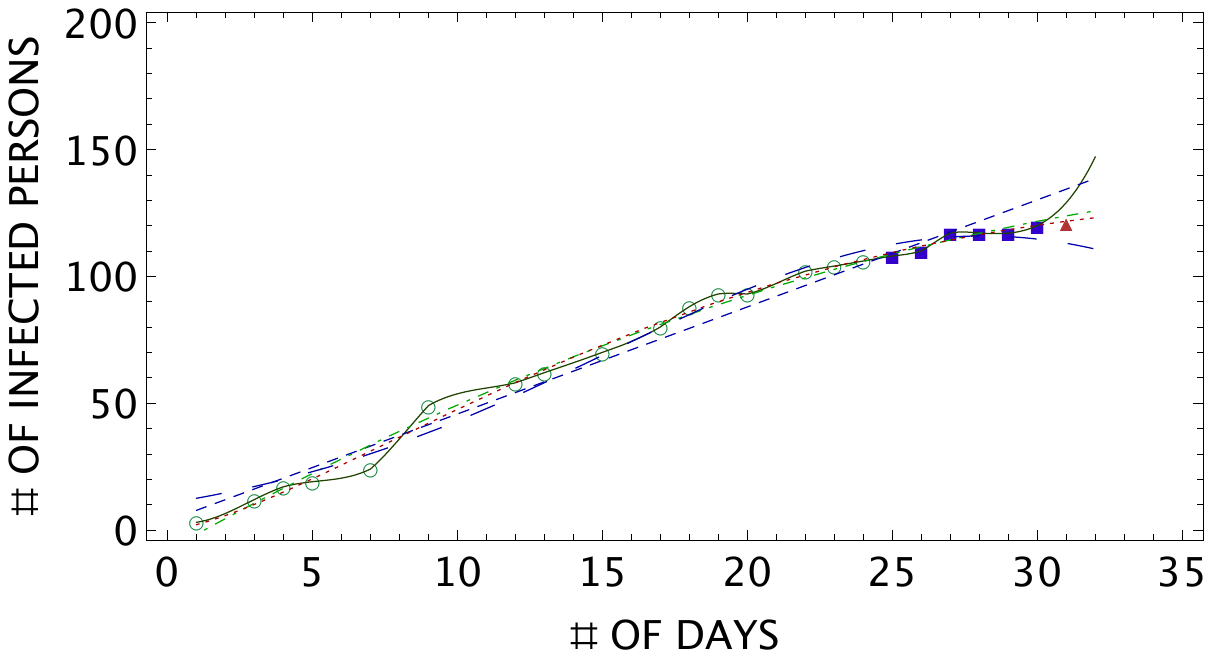}
\vspace*{-0.25cm}
\caption{\footnotesize  Data are the same as in Fig\,\ref{fig:fig3}. New data (red triangle) is not included in the fit.The continuous oliva  curve is the fit from 1 to 25th days of\,\cite{MIDI,GOOGLE} data and 25th to 30th days of data from\,\cite{MIDI,GOOGLE,OMS}. The dashed blue line is the linear fit in Eq.\,\ref{eq:linear3} while the dot-dashed green curve is the quadratic fit given in Eq.\,\ref{eq:quadratic}. The dotdashed red curve is the fit from the SIR-like model used in\,\cite{ZIFF,MALT} while the large dashed green curve is the one from a guassian-like fit used in\,\cite{TSIRONIS}.} 
\label{fig:fig4}
\end{center}
\vspace*{-0.5cm}
\end{figure}

Our results from the $\chi^2$-fit (continuous oliva line), from a linear (dashed blue line) and  a quadratic (dotdashed green curve) parametrization of the data are shown in Fig.\,\ref{fig:fig4}. 

 We shall compare our previous results with some from different approaches in Refs.\,\cite{ZIFF,MALT,MO,SARDAR,CARCIONE,HERY,GRECE1,GRECE2,TSIRONIS}, though we are aware that these models are more accurate for a large number of events.

Among these different models and for an illustration, we choose the one in\,\cite{ZIFF,MALT} using a variant of the (SIR) model\,\cite{SIR} with mixed power and exponential behaviours\,\cite{VASQUEZ} :
\vspace*{-0.2cm}
\beq
n_i(x)=ax^be^{-x/x_0}~.
\eeq
\vspace*{-0.2cm}
The data are fitted by the parameters : $a=2.2,~b=1.5$ and $x_0\approx28$ days  which we show in Fig.\,\ref{fig:fig4} (small dashed red curve). One can notice a good agreement between the quadratic polynomial fit and the one from the SIR-like model of\,\cite{ZIFF,MALT}. 

 We also fit the data using a Gaussian-like function as in\,\cite{TSIRONIS}:
\vspace*{-0.2cm}
\beq
n_i(x)=ae^{-{(x-x_0)^2\over 2\sigma^2}}~.
\eeq
\vspace*{-0.2cm}
 The result is shown in Fig.\,\ref{fig:fig3} (large dashed green curve) for a gaussian centered at $x_0\simeq 28$ days and a=115.9,  $\sigma=12.8$. 
 
 One can notice that all different models indicate a slow increase of the number of new daily infected persons which can be checked by more forthcoming data. Our Model with polynomial behaviour leads to a good fit of the data. However,  the decrease shown by the Gaussian model may be unrealistic. 
 
 \vspace*{-0.3cm}
 \subsection{Extension of the Analysis until 46 days}
 After the first version of the paper, new data based on 3700 detection tests appear (see Table\,\ref{tab:data}), where one can observe an almost stable number of new daily infected persons from 30 to 42 days which requires a reparametrization of the different models. Therefore, we extend the analysis until 46 days. The results of the analysis are summarized in  Fig.\,\ref{fig:fig5}. 

One can see from there that the mixed power- exponential model used in\,\cite{ZIFF,MALT} (dot red curve  with $a=3.74, b=1.22, x_0=44.08$) and the quadratic polynomial fit (dashed blue curve with $n_i(x)=-4.81+6.00x-0.06x^2$) give almost the same predictions, while the gaussian-like model used in\,\cite{TSIRONIS} (large dashed purple curve with $a=136.58, x_0=40.73, \sigma=21.19$) fails to describe the most recent three data. The linear fit provides a good smearing of the evolution number:
\beq
n_i(x)\simeq 20.82 + 2.98x,
\eeq
indicating an average of about 3 infected persons per day.
The best fit is obtained from the cubic polynomial :
\beq
n_i(x)\simeq -13.7+8.2x-.2x^2+1.6\times 10^{-3}x^3.
\eeq
This cubic polynomial model signals a future sharper growth while the other models predict some future stabilities or even a decrease of the number.  However, the features from these other models look unrealistic as the virus incubation period of about one month is just finished while the winter season is approaching which could be a favoured period for the virus. 
 \vspace*{-0.3cm}
\subsection{Data of the 48th day as a signal of a new phase}
We show in Fig.\,\ref{fig:fig5} the new data of the 48th day (red full circle) based on 3968 detection tests. It indicates a sudden jump compared to the previous data such that the different models successful to describe the data for the first 46 days fail here. 

However, the new data agree with the least $\chi^2$-fit method used to fit the former data. This feature may signal a new phase for the spread of the infection. It coinc\"\i des with the early partial deconfinement, the end of incubation period from contacts and the beginnng of the winter time where the UV destruction on the virus may be less efficient. 
 \vspace*{-0.3cm}
\subsection{Prediction from 48th to 58 days}
Including the new data of the 48th day, we give a prediction for the next 10 days in Fig.\,\ref{fig:fig6}.  From this analysis, we expect that the most conservative predictions for the next ten days are inside the region (purple colour) limited by the cubic polynomial (red) and least-$\chi^2$-fit (oliva) continuous curves. 

We show the collected new data for the next ten days in Fig.\,\ref{fig:58} (green full circle) which are inside the predicted region.

 \vspace*{-0.3cm}
\subsection{Fit of the 58 days data and predictions until the 67th day}
We show in Fig.\,\ref{fig:58} the analysis of the data until 58 days and the prediction until the 67th day.  Again, the region delimited by the cubic polynomial and the least-$\chi^2$-fits gives a nice prediction of the next 10 days data. 

One can notice the jump of the number of contaminated persons which indicates the 2nd phase of the pandemic from the 48th day. 

 \vspace*{-0.3cm}
\subsection{Predictions beyond the 67th day}
We show in Fig.\,\ref{fig:68} the new fit of the data before 67 days and the prediction (purple region) for the next 10 days where the cubic polynomial is parametrized as :
\beq
n_i(x)=-55.27+18.25x-0.67x^2+0.008x^3~.
\eeq
The new data (open green circles) favour the cubic polynomial behaviour. The analysis of these new data and those collected after the completion of this work is under consideration.

\section{Evolution number of cured persons}
 \subsection{Analysis  until 46 days}

We complete the paper by analyzing the evolution of the cured persons until 46 days. Fig.\,\ref{fig:cur1} shows the total number of cured persons per number of days. One can notice that the three parametrizations (linear and quadratic polynomials and the SIR-like mixed power-exponential model) provide a good description of the data with the forms :
  \bea
 n_c(x) &\simeq&4.33\times 10^{-6}x^{5.49}e^{-x/11.52}~:~~~~{\rm SIR-like ~model}\nnb\\
&\simeq&-58.85+3.49x~:~~~~~~~~~~~~~~~~~~{\rm linear},\nnb\\
&\simeq&-19.57+0.56x+0.05x^2~:~~~{\rm quadratic},\nnb\\
&\simeq&40.98-6.52x+0.30x^2-0.003x^3:\,{\rm cubic}.
 \eea

Fig.\,\ref{fig:cur2} shows the total percentage number of cured persons per number of days. The behaviour is similar to the one in 
 Fig.\,\ref{fig:cur1} and is parametrized as :
 \bea
 n_c\%(x) &\simeq&1.56\times 10^{-6}x^{6.00}e^{-x/8.61}~:~~~{\rm SIR-like ~model}\nnb\\
&\simeq&-37.37+2.46x~:~~~~~~~~~~~~~~~~~~{\rm linear},\nnb\\
&\simeq&-33.60+2.18x+0.005x^2~:~~~{\rm quadratic},\nnb\\
&\simeq&41.27-6.57x+0.32x^2-0.004x^3:\,{\rm cubic},\nnb\\
 \eea

From this analysis, one can deduce from the linear fit an average number of about 3 cured persons per day which correspond to about 2.4\% of the total number of declared infected persons. 

 One can observe the relative good performance of the medical care (2.9-3.5) cured persons per day compared to (4-5) infected persons per day and that no deceased persons have been officially declared since the beginning of the pandemic.
 \vspace*{-0.3cm}
 \subsection{Analysis  of the 48th day}
The 48th day corresponds to a jump for the number of new daily infected persons which we consider as the beginning of the new phase of the infected persons.

One can see from Fig.\,\ref{fig:cur1} that the new data on the 48th day on the total number of cured persons can be better predicted within the error by the different models used to analyze the data below 46 days except the quadratic polynomial.  

However, for the percentage number of cured persons, only the least $\chi^2$-fit can give a good prediction of the 48th day data as can be seen in Fig.\,\ref{fig:cur2}.  

These features indicate that with the new phase, the parametrization of the models should be updated.  The sudden growth of the number of new daily infected persons also indicates the relative decrease of the care performance.
 \vspace*{-0.3cm}
 \subsection{Analysis until the 67th day}

We show in Fig.\,\ref{fig:43} the evolution of the absolute number of cured persons which is approximately linear with a slope of about 3 persons cured per day. The percentage of the cured over the contaminated numbers is shown in  Fig.\,\ref{fig:53} which is approximately quadratic and has a peak around 42-46 days. The decrease of the curve above 46 days indicates the unability of the hospital equipment and drugs to face the 2nd phase of the pandemic. It also indicates that the new CVO artemisia based tisane  publicly announced by the Madagascar President on 29th april is not at all efficient  until  the 67th day of pandemic.

\section{COVID-19 and Confinement for Madagascar}
\vspace*{-0.15cm}
 \subsection{Political decisions}
As already mentioned in the introduction, in order to limit the spread of this dangerous virus which propagates via social contacts (coughing , saliva droplets,..) and waiting for a new efficient medical drugs, different countries have decided to confine the population and have asked the persons to strictly respect some barriers rules. About 3 billions of persons i.e half of the humanity are now / have been confined. 

 \vspace*{-0.3cm}
 \subsection{Implementation}
However, the implementation of confinement in developing countries is far to be achieved  due to poverty and to the lack of education of the majority of the population\,:

Due to poverty and to the bad organization of the society, most of persons have to work for finding what to eat day by day due to the  informal forms of the trade, business and economy. 

In addition,  the accompanying help measures taken by the government are  insufficient while the managements of some  international funds and donations are not transparent. 

Middle class persons also suffers as they are not rich enough to be autonomous and not too poor to receive any help from the state. 

Due to the lack of education, most of the people are irresponsable and are not aware about the dangerous effect of the virus. Then, they do not see the importance of confinement and do not feel obliged to respect the barriers measures and the confinement. 

\vspace*{-0.25cm}
\section{Worldwide Impacts of COVID-19 and Confinement}

More generally and as a consequence, these unprecedented  pandemic and confinement security measures have large impacts for human beings :
\vspace*{-0.25cm}
\subsection{Social organisation}
\b Most of us learn about teleworks, indoor at outdoor houseworks. 

\b We re-discover the importance of a family and of the  tradition 

\b We re-discover ourselves from our concentration and meditation.

\b We see the usefulness of solidarity.

\b We see the values of health personals, researchers, teachers, educators. firefighters,...and in general the human values. 

\b Urban exodus in developing countries such as Madagascar are observed.

\b However, the pandemic might enhance the social class inequalities like the increase of the difference between poor and rich peoples, the re-disappearance of middle class, mentioned earlier for Madagascar, which is the lungs of developments. 

\vspace*{-0.25cm}
\subsection{Environments}
\b Nature takes back its rights : returns of animals near cities and cetaceans near the coasts, returns of insects, birds,...). 

\b  Air pollution is decreasing due to a minimal road traffic.  Megacities (Beijing, New-Delhi, New-York, Parish,...) recover an improved atmospheric air.

\b Cities become less noisy.

\b ...

\vspace*{-0.25cm}
\subsection{Economy}

\b Globalization is suffering.

\b Reduced size and local markets are developing.

\b  Small producers are reorganizing.

\b On-line sales and drive markets are developing. 

\b Delocalisations of manufacturers and factories are questioned.

\b Each country is looking for an independent and self-sufficient economy which may lead to a protectionism. 

\vspace*{-0.25cm}
\subsection{Politics}
\b COVID-19 has pressed the leaders of each country to revisit the orientation of their politics  to the most useful ones for the population namely\,:

\hspace*{0.5cm}-- Health,
  
\hspace*{0.5cm}-- Research and Education, 
  
\hspace*{0.5cm}-- Foods and different ways of Consuming,  
  
\hspace*{0.5cm}-- Protections of the Environments.
  
 \b COVID-19 has stimulated the search of an autonomous country which, if not done carefully,  may lead to an ultra-nationalism and to a withdrawl into oneself (closing of boarders,...)

\vspace*{-0.25cm}
\section{Summary and Conclusions}
We have studied the first month spread of COVID-19 in Madagascar using standard least $\chi^2$ fit approach. We found that, for this first month, the spread of the virus per day increases almost linearly/quadratically for about (4-5) persons per day. A comparison of our results with the ones from some other SIR-like models is shown in Fig.\,\ref{fig:fig4}. 

\begin{figure}[hbt]
\vspace*{-0.25cm}
\begin{center}
\includegraphics[width=8.cm]{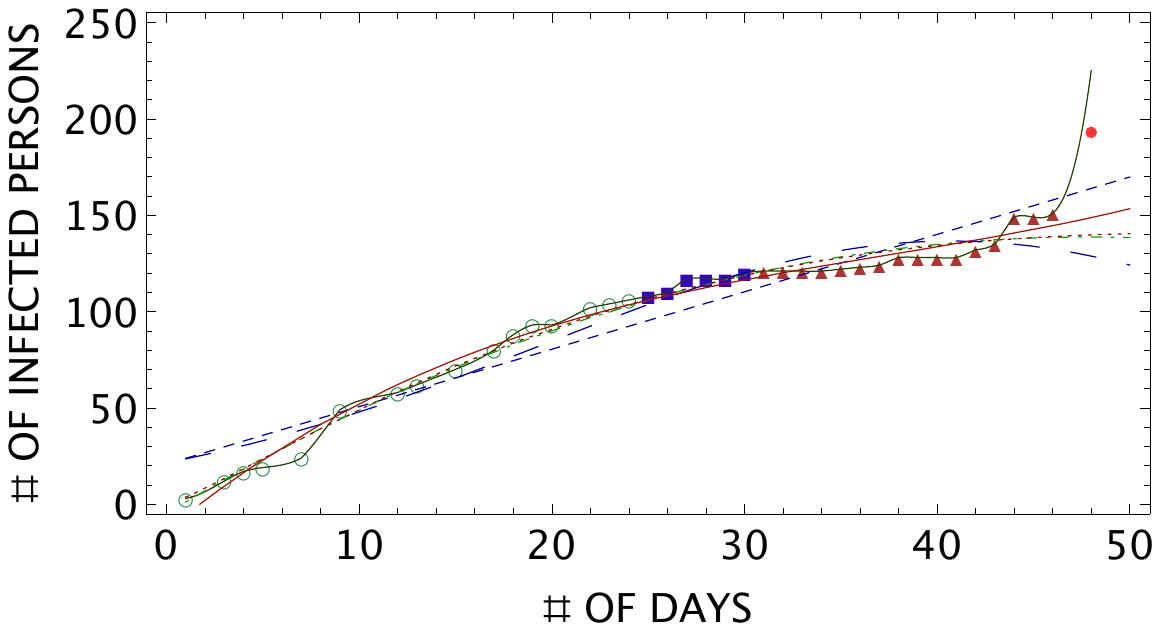}
\vspace*{-0.25cm}
\caption{\footnotesize  Previous data are the same as in Fig\,\ref{fig:fig4}. New data (dark red triangles) are added in the analysis.The dashed blue line is the linear fit in Eq.\,\ref{eq:linear3}. The dot-dashed green curve is the quadratic fit given in Eq.\,\ref{eq:quadratic}. The full red curve is the cubic fit. The dotdashed red curve is the fit from the SI-like model used in\,\cite{ZIFF,MALT} while the long dashed blue curve is the one from a guassian-like fit used in\,\cite{TSIRONIS}. The continuous  oliva curve is the one from of least $\chi^2$-fit. The  red point is the new data on 7th may not included in the fit.} 
\label{fig:fig5}
\end{center}
\vspace*{-0.5cm}
\end{figure}

\begin{figure}[hbt]
\vspace*{-0.25cm}
\begin{center}
\includegraphics[width=8.cm]{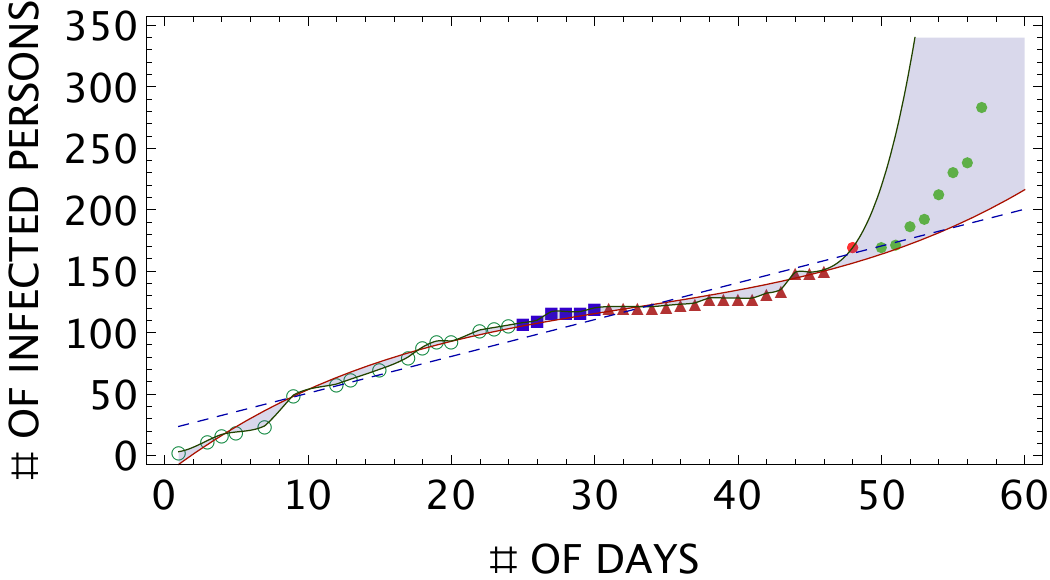}
\vspace*{-0.25cm}
\caption{\footnotesize  Data are the same as in Fig\,\ref{fig:fig5}. New fit including the new data 169 (red point) on 7th may (48th day). The oliva curve is the $\chi^2$-fit. The continuous red curve is the cubic polynomial. The dashed line is the linear fit. The pink region is the prediction where we show the new data (not used in the fit) collected until the 58th day.} 
\label{fig:fig6}
\end{center}
\vspace*{-0.5cm}
\end{figure}

\begin{figure}[hbt]
\vspace*{-0.25cm}
\begin{center}
\includegraphics[width=8.cm]{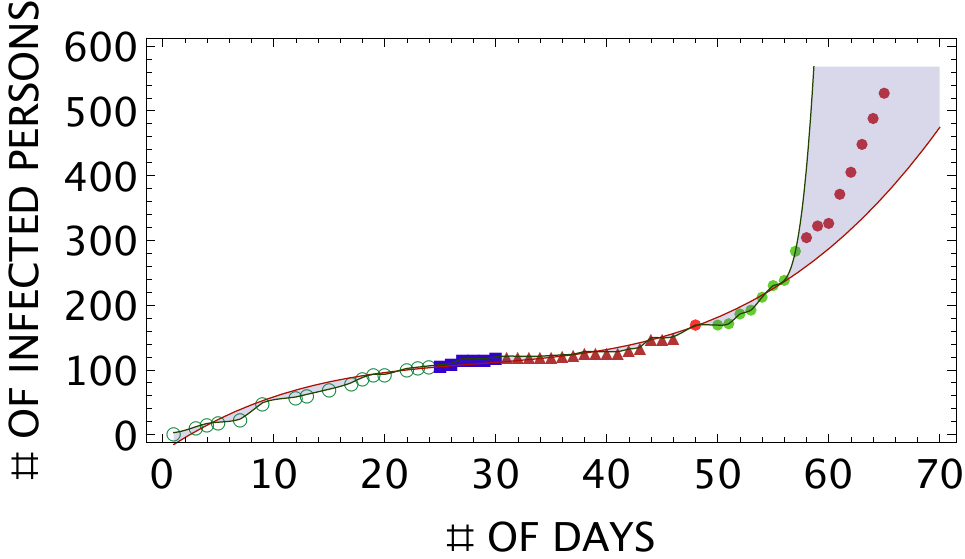}
\vspace*{-0.25cm}
\caption{\footnotesize  New fit including the new data (green full circle). The oliva curve is the $\chi^2$-fit. The continuous red curve is the cubic polynomial. The purple region is the prediction where we show the new data (not used in the fit) collected until the 67th day.} 
\label{fig:58}
\end{center}
\vspace*{-0.5cm}
\end{figure}

\begin{figure}[hbt]
\vspace*{-0.25cm}
\begin{center}
\includegraphics[width=8.cm]{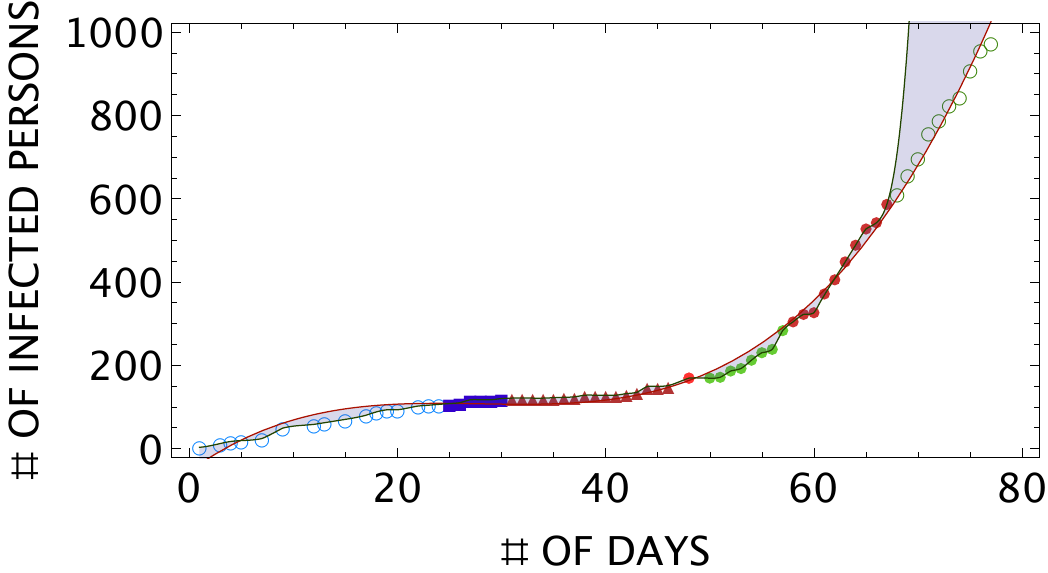}
\vspace*{-0.25cm}
\caption{\footnotesize  New fit including data between 59 to 67 days (red full circle). The oliva curve is the $\chi^2$-fit. The continuous red curve is the cubic polynomial. The purple region is the expected prediction. The green open circles are new data from 68th to 77th days not included in the fit. } 
\label{fig:68}
\end{center}
\vspace*{-0.5cm}
\end{figure}

\begin{figure}[hbt]
\vspace*{-0.25cm}
\begin{center}
\includegraphics[width=8.cm]{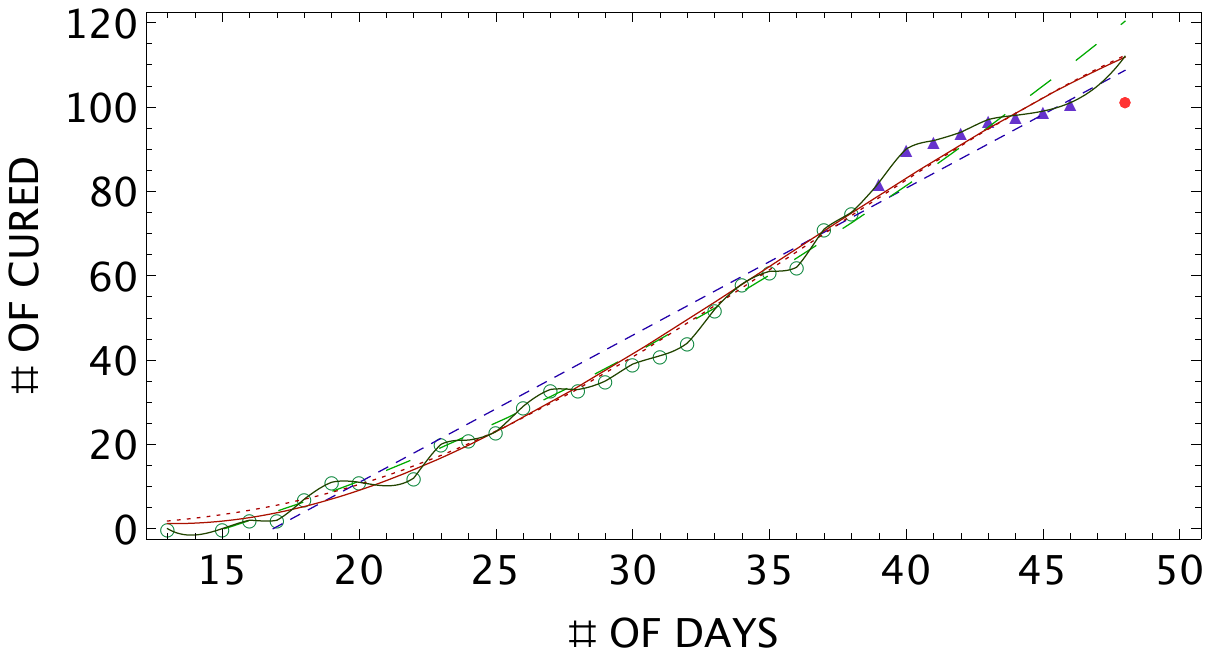}
\vspace*{-0.25cm}
\caption{\footnotesize  Comparison of different model predictions using the data in Table\,\ref{tab:data} until 46 days. The dashed blue line is a linear fit . The long dashed green curve is a quadratic fit in Eq.\,\ref{eq:quadratic}. The 
dot red curve for the SI-like model in\,\cite{ZIFF,MALT} and the continuous red curve for a cubic polynomial fit are almost superimposed. The green continuous curve is the $\chi^2$-fit. Blue triangles are new data until 46 days. The red point is data of the 48th day not included in the fit.} 
\label{fig:cur1}
\end{center}
\end{figure}

\begin{figure}[hbt]
\vspace*{-0.25cm}
\begin{center}
\includegraphics[width=8.cm]{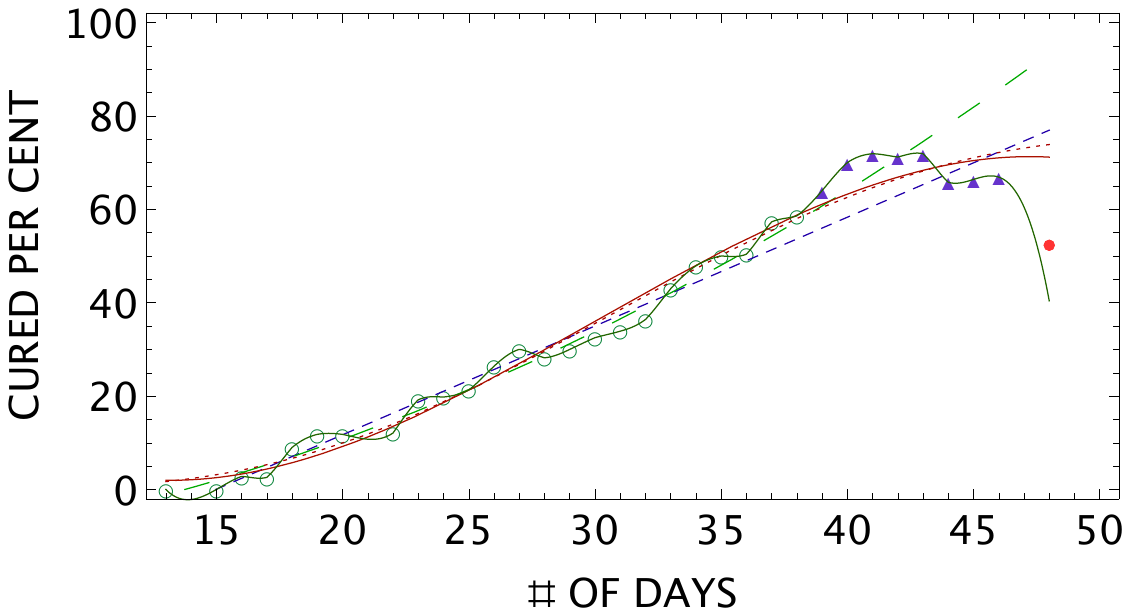}
\vspace*{-0.25cm}
\caption{\footnotesize  The same as in Fig.\,\ref{fig:cur1} but for the per cent number of cured persons.
The data come fromTable\,\ref{tab:data}. Green circles are data until 38 days. Blue triangles are new data until 46 days. The red point is data from the 48th day not included in the fit.} 
\label{fig:cur2}
\end{center}
\vspace*{-0.5cm}
\end{figure}

\begin{figure}[hbt]
\vspace*{-0.25cm}
\begin{center}
\includegraphics[width=8.cm]{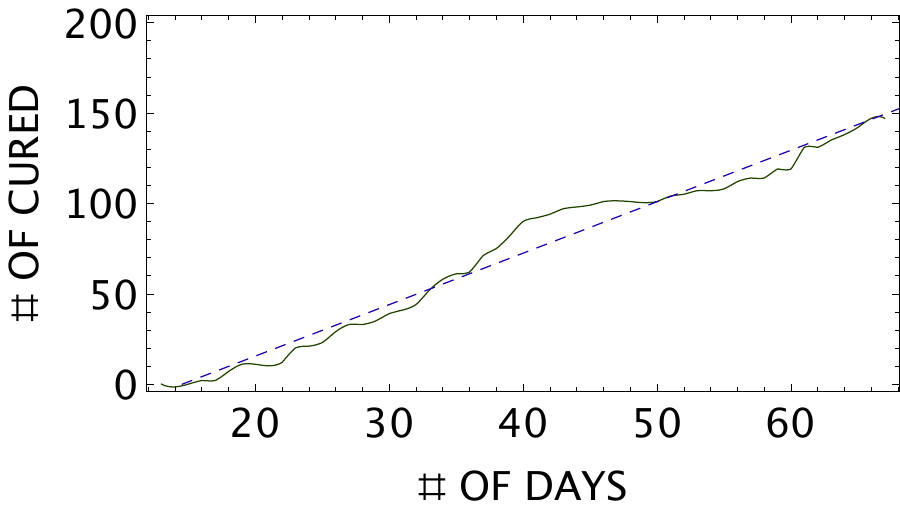}
\vspace*{-0.25cm}
\caption{\footnotesize  Absolute number of cured persons for 67 days. The dashed curve is a linear fit.} 
\label{fig:43}
\end{center}
\vspace*{-0.5cm}
\end{figure}

\begin{figure}[hbt]
\vspace*{-0.25cm}
\begin{center}
\includegraphics[width=8.cm]{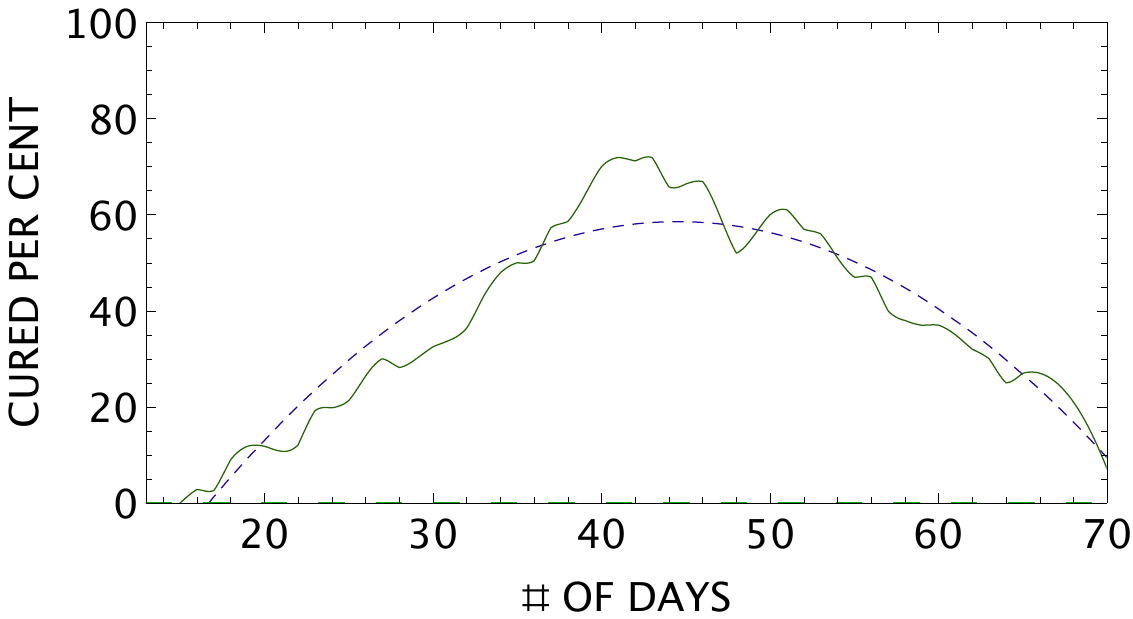}
\vspace*{-0.25cm}
\caption{\footnotesize  The same as in Fig.\,\ref{fig:43} but for the per cent number of cured persons. The dashed curve is a quadratic fit.} 
\label{fig:53}
\end{center}
\vspace*{-0.5cm}
\end{figure}

The inclusion of the recent data from 28 to 46 days indicates new features which are essentially due to the plateau from 30 to 40 days (see Fig.\,\ref{fig:fig5}) and requires a new parametrization of the data. Indeed, these additional  data favour a cubic polynomial fit instead of the quadratic one and of the models used in\,\cite{ZIFF,MALT,TSIRONIS}. 

These early data do not indicate  any sign of an exponential growth of the number of new daily infected persons though the cubic polynomial signals a stronger increase in the near future. 

Indeed, more recent data collected on the 48th days show a sudden jump  which are not explained by the previous models except the least $\chi^2$ fit as shown in Fig.\,\ref{fig:fig5}. This sudden jump signals a new phase on the spread of the virus which may coincide with the end of the incubation period (contact case), the arrival of the winter period which the virus may like and with the beginning of the partial deconfinement. 

We show in Fig.\,\ref{fig:fig6} the new fit including this new data on the 48th day. The prediction for the next days are given by the purple region limited by the $\chi^2$-fit (oliva curve) and the cubic polynomial (red curve).
Forthcoming new data from the 48th to the 58th days are shown inside this region. 

A similar analysis is done in Fig.\,\ref{fig:58} using the data until the 58th day and the predicted purple region where the new data from the 59th to the 67th days are shown (full red circle).

We note that, for the first month, the number of new daily infected persons are relatively low, which can be due to the reduced number of detection tests. Another factor could be that the intense UV in the country (autumn season) during this first month may have partially neutralized the virus effects. 

One should note that we have done a simple least square fit and polynomial parametrization to understand the spread of the pandemic during these first months which we have compared with two SIR-models used in\,\cite{ZIFF,MALT}. We expect that our results will be useful for new model buildings. 

We have not discussed the basic reproductive number $R_0$-coefficient due to the insufficient pandemic tests in Madagascar for controlling more accurately this effect. An attempt to estimate $R_0$ for the week of 26 june to 2 july indicates that $R_0$  ranges from 1.2 to 1.7 in Antananarivo\,\cite{HERY2} but the result can vary strongly from one region to another.

We have also analyzed the number of cured persons where we note that its behaviour is almost linear with a slope of about 3 persons per day. The linear/quadratic polynomial fit and the SI-like model with mixed power-exponential behaviours fit quite well the data before the 40th days but, with the inclusion of the new data from 38 to 46 days,  the whole data are better fitted  by a cubic polynomial like in the case of infected persons. 

One has also found that the percentage number of person cured corresponding to the new data beyond 48 days are not reproduced by the different models discussed above
but only by the least $\chi^2$-approach which signals a relative decrease of the percentage care as a consequence of the sudden increase number of new daily infected persons. 

It is remarkable to note that the total number of about (2.9-3.5) officially cured persons per day relative to the number (4-5) of infected persons for the first month is a good performance despite the poor equipment of the hospitals.  Then, one wonders if these care techniques are based on modern chemical drugs or complemented by some endemic medicinal plants ? More explanation, on the technical cares used to get these successful results,  is greatly appreciated.

However, the duration of the previous success is short as the percentage of cured relative to the contaminated persons has decreased drastically after the 42-46 days signaling the inefficiency or/and the limit of the hospital care after the 48th day jump of the number of contaminated persons. 

During the period of our study until the 67th day of pandemic, there is also no indication of the efficiency of the new artemisia CVO based tisane mixed with some endemic plants (ravintsara,...) distributed from 29th april which is claimed by the President of Madagascar to prevent and cure the COVID-19 pandemic. 

We have also noticed the relatively low number of tested persons which is around  9681 persons on 27th may 2020 compared to the total number 27 millions of population where 1.6 millions live in the capital Antananarivo. A large effort for increasing this number of detection tests, its origin and the kind of samples  is required to have a more realistic basis for a more robust scientific analysis.

More transparency on the uses of the international donations for the hosptial equipments, materials for detection tests, barrier material (masks, hydroalcoholic gel,...), the help to the poor peoples is continuously requested. 

We have shortly reviewed the general Worldwide impacts of COVID-19 which  has demonstrated the weakness of the current global system. This feature might announce a  change towards a new model of society and economy and for a new form of political decisions in the near future. 

\section*{Note added} During the submission of the paper, the spread of the pandemic in Madagascar has drastically changed. At present the number of contamined persons is about 100 times the one studied in this paper and it seems to behave exponentially where a peak has been eventually reached.We plan to study this new feature in a future work.

\section*{Acknowledgements}
It is a pleasure to thank S. Maltezos, R. Narison and R. Ziff for reading the manuscript and for some communications.

This research has not received any dedicated funds and sponsoring. It is issued from a personal initiative without any influence from private and public organizations.

\vspace*{0.5cm}

{\scriptsize
\begin{table}[hbt]
\vspace*{-0.25cm}
 \caption{The top ten countries having the highest numbers of contaminated persons and above 100 000 persons on 08/05/2020. The data come from\,\cite{GOOGLE}.}  
\setlength{\tabcolsep}{0.9pc}
    {\small
  \begin{tabular}{clrr}
&\\
\hline
\hline
Rank &Country &Absolute \# & \# per million  \\
\hline
1&USA & 1 289 235 & 3 912\\
2&Spain & 221 447 & 4 702\\
3&Italy & 215 858&3 583 \\
4&United Kingdom & 206 715 &3 112\\
5&Russia &177 160 & 1 207\\
6&Germany&169 430 & 2 038 \\
7&France & 137 779 & 2 054 \\
8&Brazil & 135 773 & 642 \\
9& Turkish & 133 721 & 1 608 \\
10&Iran & 103 135 & 1 238 \\
\hline\hline
\end{tabular}
}
\label{tab:countries}
\end{table}
} 

{\scriptsize
\begin{table}[hbt]
 \caption{The top ten countries having the relative highest numbers of contaminated persons per million of population on 08/05/2020. The data come from\,\cite{GOOGLE}.}  
\setlength{\tabcolsep}{1 pc}
    {\small
  \begin{tabular}{clrr}
&\\
\hline
\hline
Rank &Country & \# per million &Absolute \#  \\
\hline
1&San Marino & 18 526&\,\, 622\\
2&Andorra &\,\, 9 698 &\,\, 752\\
3&Qatar &\,\,6 876 &\,\,18 890   \\
4&Luxembourg &\,\,6 286 &3 859 \\
5&Island & \,\,4 944& 1 801\\
5&Spain & \,\,4 702& 221 447\\
6&Irland & \,\,4 548& 22 385\\
7&Belgium & \,\,4 462& 51 420\\
8&Gibraltar &\,\,4 273 &\,\, 144\\
9&Guernesey &\,\,4 013 &252  \\
10&Singapour &\,\,3 671 &20 939\\ 
\hline\hline
\end{tabular}
}
\label{tab:countries2}
\end{table}
} 
{\scriptsize
\begin{table}[hbt]
 \caption{Numbers of Contaminated Persons in Madagascar at different dates. The data from 21/03/ to 15/04/ have been communicated by the national press\,\cite{MIDI} and TV\,\cite{TVM} issued from the WHO / OMS agency in Madagascar\,\cite{CCO} and from Google\,\cite{GOOGLE}. Data compiled by WHO / OMS are from\,\cite{OMS}. The added quoted errors are an estimate of about (10-20)\% systematics which can be an underestimate. The 2 last columns are the number and percent of cured persons.}  
\setlength{\tabcolsep}{0.6pc}
    {\scriptsize
\begin{tabular}{ccccccc}
&\\
\hline
\hline
Date &\# of Days &\# of Infected  && \# of Cured&\% of Cured \\
\hline
&&Ref\,\cite{MIDI,TVM,CCO}&&Ref\,\cite{TVM,CCO}&Ref\,\cite{TVM,CCO}\\
\cline{3-4}
21/03 &1& 3(1)&&0&0 \\
23/03 &3&12(3)&&0&0\\
24/03 &4& 17(4)&&0&0 \\
25/03&5& 19(5)&&0&0\\
27/03&7&24(5)&&0&0\\
29/03&9&49(5)&&0&0\\
01/04&12 &58(6)&&0&0\\
02/04 &13& 62(6)&&0&0\\
04/04 &15& 70(7) &&0&0\\
05/04&16&72&&2&2.8\\
\hline
&& \multicolumn{1}{c}{Ref\,\cite{MIDI,TVM,CCO,GOOGLE}}& \multicolumn{1}{c}{Ref.\,\cite{OMS}} \\
\cline{3-4}
06/04 &17& 80&77&2&2.6\\
07/04 &18& 88&77&7&9.0 \\
08/04 & 19 &93&92&11&11.8\\
09/04&20&93&93&11&11.8\\
11/04&22&102 &95&12&12.1\\
12/04&23&104&104&20&19.2\\
13/04&24&106&106&21&19.8 \\
\hline
14/04&25&108&--&23&21.3\\
15/04&26&110&--&29&26.4 \\
16/04&27&111&110&33&30.0\\
17/04&28&--&117&33&28.2\\
18/04&29&--&117&35&30.0\\
19/04&30&120&--&39&32.5\\
\hline
20/04&31&121&121&41&33.9\\
21/04&32&121&121&44&36.4\\
22/04&33&121&121&52&43.0\\
23/04&34&121&121&58&47.9\\
24/04&35&122&122&61&50.0\\
25/04&36&123&123&62&50.4\\
26/04&37&124&--&71&57.3\\
27/04&38&128&--&75&58.6\\
28/04&39&128&--&82&64\\
29/04&40&128&--&90&70\\
30/04&41&128&--&92&71.9\\
\hline
01/05&42&132&--&94&71.2\\
02/05&43&135&--&97&71.9\\
03/05&44&149&--&98&65.8\\
04/05&45&149&--&99&66.4\\
05/05&46&151&--&101&66.9\\
\hline
07/05&48&169 &--&101&59.8\\
09/05&50&169&--&101&60\\
10/05&51&171&--&104&61\\
11/05&52&186&--&105&57\\
12/05&53&192&--&107&56\\
13/05&54&212&--&107&51\\
14/05&55&230&--&108&47\\
15/05&56&238&--&112&47\\
16/05&57&283&--&114&40\\
17/05&58&304&--&114&38\\
\hline
18/05&59&322&--&119&37\\
19/05&60&326&--&119&37\\
20/05&61&371&--&131&35\\
21/05&62&405&--&131&32\\
22/05&63&448&--&135&30\\
23/05&64&488&--&138&25\\
24/05&65&527&--&142&27\\
25/05&66&542&--&147&27\\
26/05&67&586&--&147&25\\
\hline
27/05 &68& 612&--&151&24.7 \\
28/05 &69& 656&--&154&23.48\\
29/05 &70& 698&--&164&23.5\\
30/05 &71& 758&--&165&21.8 \\
31/05&72&790&--&168&21.3\\
01/06 &73& 826&--&174&21.07\\
02/06 &74& 845&--&185&21.89\\
03/06 &75& 908&--&195&21.47\\
04/06 &76& 957&--&200&20.89\\
05/06 &77& 975&--&201&20.62 \\
\hline\hline
\end{tabular}
}
\label{tab:data}
\end{table}
} 

\begin{thebibliography}{999}
\bibitem{PASTEUR} S. Van der Werf et al. (Institut Pasteur, CNRS, Univ. Paris VII),  Fascicule de Brevet Europeen EP1 694 829 B1 (2006).
\bibitem{MIDI} Midi-Madagasikara newpaper daily press.
\bibitem{TVM}Television Malagasy (TVM) daily news.
\bibitem{CCO}Centre de Commandement Operationnel (CCO) Ivato-Antananarivo, WHO's national agency.
\bibitem{GOOGLE} Google : https://google.com/covid19-map.
\bibitem{OMS}WHO/OMS compilation : https://www.who.int/emergencies/diseases/
novel-coronavirus-2019/situation-reports. 
\bibitem{MATH} Wolfram  research foundation. 
\bibitem{SANTE} Sant\'e Magazine, article of 18/03/2020.
\bibitem{ARXIV} https://arxiv.org/search/?query=COVID-19. 
\bibitem{SIR}W. O. Kermack, A. G. McKendrick, {\it Proc. Royal Soc. London}, {Vol. 115, Issue 772} (1927) 700.
\bibitem{ZIFF} A. L. Ziff and R.M. Ziff, medRxiv 2020.02.16.20023820. 
\bibitem{MALT}S. Maltezos, arXiv:2004.05992 [physics.soc-ph].
\bibitem{MLI}M. Li et al., arXiv:2002.09199 .
\bibitem{ZLIU}Z. Liu et al., arXiv:2002.12298 [q-bio.PE].
\bibitem{MO}B. Mo et al., arXiv:2004.04602 [physics.soc-ph].
\bibitem{SARDAR}T. Sardar, S.S. Nadim and J. Chattopathyya, arXiv:2004.03487 [q-bio.PE].
\bibitem{SAVI}P. V. Savi, M. A. Savi and B. Borges, rXiv:2004.03495 [q-bio.PE].
\bibitem{CARCIONE}J. M. Carcione et al., arXiv:2004.03575 [q-bio.PE].
\bibitem{HERY}J. Dehning et al., arXiv:2004.01105 [q-bio.PE].
\bibitem{HERY2}H. Ratsimbarison, https://github.com/herysedra/ady\_cov/ (2020). 
\bibitem{GRECE1}C. Anastassopoulou et al.,   https://doi.org/10.1371/journal.pone.0230405.
\bibitem{GRECE2}C. Siettos and L. Russo, https://www.tandfonline.com/loi/kvir20.
\bibitem{TSIRONIS}G. D. Barmparis and G. P. Tsironis, DOI: 10.1016/j.chaos.2020.109842 (arXiv:2003.14334 [q-bio.PE]). 	.
\bibitem{VASQUEZ} A. Vasquez, {\it Phys. Rev. Lett.} 96.038702 (2006).
\bibitem{SNBm}S. Narison, {\it R\'eflexions sur Madagascar, clin d'oeil sur le pass\'e et vision vers le futur}, ISBN 97811796561982, amazon.com (2018).
\end{thebibliography}
\end{document}